\documentclass[12pt]{article}
\pdfoutput=1

\usepackage{jheppub}
\usepackage{afterpage}
\usepackage[utf8]{inputenc} 
\usepackage{bmpsize}

\usepackage{bm}
\usepackage{amsmath}
\usepackage{eufrak}

\title{Real Time Quantum Gravity Dynamics from Classical Statistical Yang-Mills Simulations}

\author{Masanori Hanada$^{1}$ and}
\affiliation{$^1$ Department of Physics, University of Colorado, Boulder, Colorado 80309, USA}

\emailAdd{masanori.hanada@colorado.edu}
\author{Paul Romatschke$^{1,2}$}
\affiliation{$^2$ Center for Theory of Quantum Matter, University of Colorado, Boulder, Colorado 80309, USA}
\emailAdd{paul.romatschke@colorado.edu}

\abstract{We perform microcanonical classical statistical lattice simulations of SU($N$) Yang-Mills theory with eight scalars on a circle.
  Measuring the eigenvalue distribution of the spatial Wilson loop  we find two distinct phases depending on the total energy and circle radius, which we tentatively interpret as corresponding to black hole and black string phases in a dual gravity picture. We proceed to study quenches by first preparing the system in one phase, rapidly changing the total energy, and monitoring the real-time system response. We observe that the system relaxes to the equilibrium phase corresponding to the new energy, in the process exhibiting characteristic damped oscillations. We interpret this as the topology change from black hole to black string configurations, with damped oscillations corresponding to quasi-normal mode ringing of the black hole/black string final state. This would suggest that $\alpha^\prime$ corrections alone can resolve the singularity associated with the topology change.  We extract the real and imaginary part of the lowest-lying presumptive quasinormal mode as a function of energy and $N$.
  }

\begin{document}
\maketitle
\section{Introduction}

General relativity breaks down when curvature singularities appear. 
How these singularities are resolved in a consistent extension of general relativity is a very important issue.

The topology change from black hole (BH)  to black string (BS) \cite{Gregory:1993vy} is an interesting physical process for which the singularity resolution is essential: although rich dynamics beyond linear perturbation theory is expected based on results in numerical relativity (see e.g. \cite{Choptuik:2003qd}), 
the change in topology requires physics beyond general relativity.

In string theory, gauge/gravity duality \cite{Maldacena:1997re} provides us with a well-defined description of this BH/BS transition in terms of a gauge theory dual\cite{Aharony:2004ig,Aharony:2005ew} (see also \cite{Susskind:1997dr,Barbon:1998cr,Li:1998jy,Martinec:1998ja}). To review the main aspects of this description, let us consider the example of two-dimensional maximally supersymmetric SU$(N)$ Yang-Mills (2d maximal SYM) compactified on spatial circle of circumference $r_x$, which is conjectured to possess a dual string theory description. In fact, this gauge theory has \textit{two} dual description, one being type IIB string theory on ${\mathbb R}^{1,8}\times$S$^1$ with $N$ D1-branes wrapped on the circle, and the other one being type IIA string theory on a circle of circumference $r_x^\prime=\frac{(2 \pi)^2 \alpha^\prime}{r_x}$ with $N$ D0-branes (known as T-dual). 

\begin{figure}[t]
  \begin{center}
     \rotatebox{0}{
   \includegraphics[width=.9\linewidth]{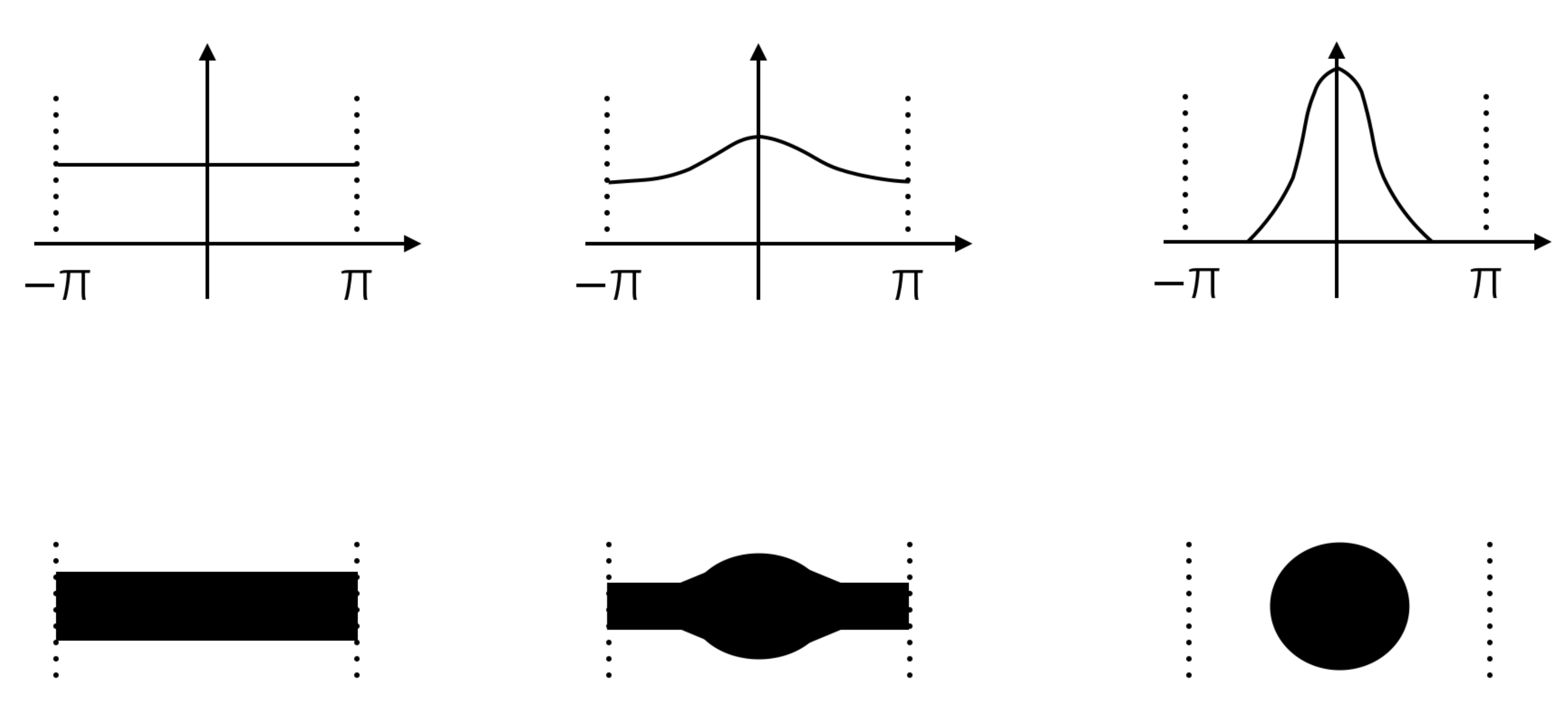}}
  \end{center}
  \caption{The conjectured correspondence between the distribution of the phases of Wilson loop eigenvalues (top row) and the topology of the black hole/black string dual configurations (bottom row), with black hole, uniform black string and wavy string configurations corresponding to localized (``gapped''), uniform and nonuniform (``ungapped'') phase distributions, respectively. Dotted lines indicate periodic boundary conditions on the circle. Figure adapted from Ref.~\cite{Hanada:2016qbz}. 
  }\label{fig:BH-BS}
\end{figure}

In terms of the 2d maximal SYM gauge theory description, the information about the individual D0-branes on the spatial circle is encoded in the phases of the eigenvalue distribution of the Wilson line winding on circle (referred to as Wilson loop in the following). Depending on the distribution of D0-branes along the circle, there can be various phases, such as a black hole phase (corresponding to a localized, or ``gapped'', distribution of Wilson loop phases), and a black string phase (corresponding to an ``ungapped'' distribution of Wilson loop phases). In the gravity dual of 2d maximal SYM gauge theory, one can further distinguish between a uniform black string phase (corresponding to a uniform distribution of Wilson loop phases), and a wavy string phase (corresponding to a nonuniform distribution of Wilson loop phases); see Fig.~\ref{fig:BH-BS}. Note that these phases have been explicitly been constructed in Ref.~\cite{Dias:2017uyv}.

\begin{figure}[htbp]

    \includegraphics[width=.49\linewidth]{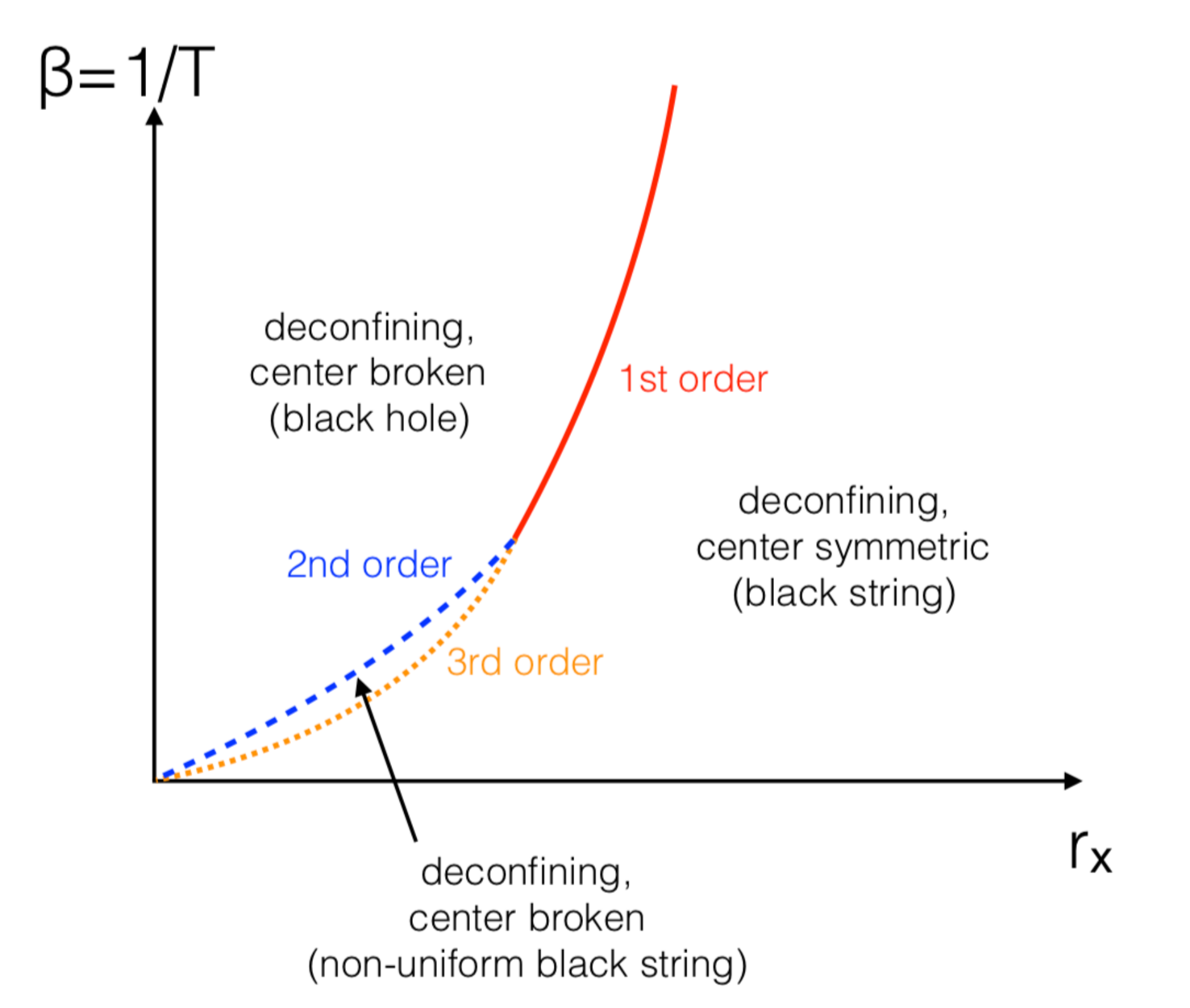}
    \includegraphics[trim=0 -1.5cm 0 0, width=.49\linewidth]{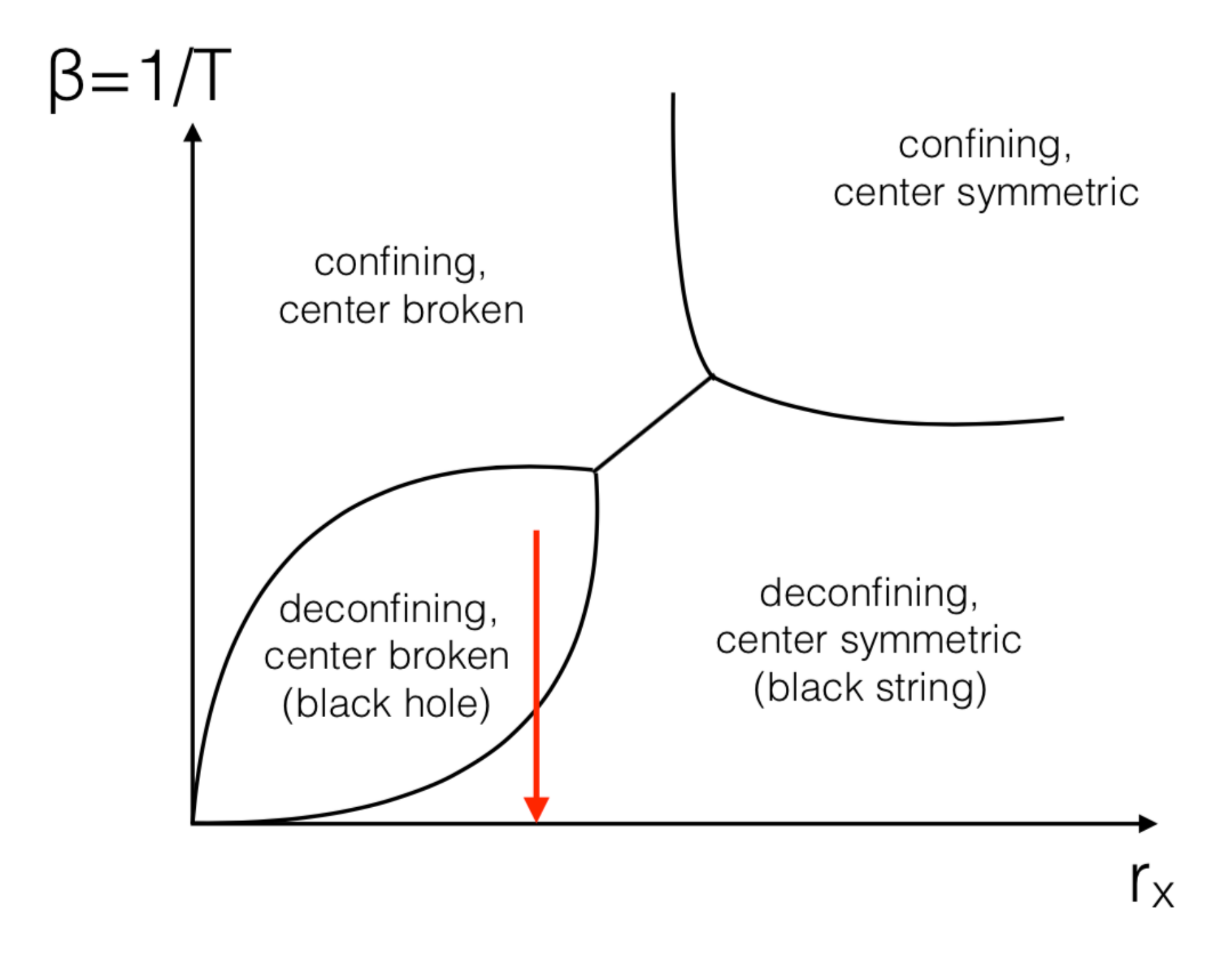}
  \caption{Conjectured equilibrium phase diagram for the gravity dual to 2d maximal SYM (left) and purely bosonic Yang-Mills with 8 scalars (right) as a function of the length of the spatial ($r_x$) and temporal ($\beta$, inverse temperature) circles, respectively. The arrow on the rhs panel indicates applicability of numerical simulations in this work. 
}\label{fig:PhaseDiagram}
\end{figure}

Equilibrium properties of the 2d SYM gauge theory can be studied by using lattice Monte Carlo simulations, which are non-trivial to set up ~\cite{Kaplan:2002wv,Cohen:2003xe,Cohen:2003qw,Kaplan:2005ta,Sugino:2003yb,Sugino:2004qd,Sugino:2004uv,Catterall:2004np,Catterall:2005fd,DAdda:2005zk,Suzuki:2005dx,Hanada:2017gqc}, but seem to be able to offer fully non-perturbative insights \cite{Hanada:2009hq,Hanada:2010qg,Giguere:2015cga,August:2018esp,Catterall:2017lub}. As a result of numerical studies in 2d maximal SYM \cite{Catterall:2010fx,Kadoh:2017mcj,Catterall:2017lub} combined with numerical and analytical studies of the thermodynamic properties of D-branes \cite{Aharony:2004ig,Aharony:2005ew,Kawahara:2007fn,Hanada:2016qbz,Mandal:2009vz}, the conjectured equilibrium phase diagram sketched in Fig.~\ref{fig:PhaseDiagram} has started to emerge.\footnote{
Note added in April 2020: Recent study \cite{Bergner:2019rca} strongly suggests that this conjecture is wrong; the transition is likely to be of first order everywhere, and the non-uniform black string phase is not thermodynamically dominant in the canonical ensemble. 
Still, this phase is dominant in the microcanonical ensemble at the intermediate energy scale, and the results of the classical simulations presented in this paper, 
which corresponds to the microcanonical ensemble, can capture the properties of this phase, without any essential change.
} 
Specifically, at low temperature $T$ (large temporal radius $\beta=T^{-1}$), two phases of localized and uniform eigenvalue phase distributions at small and large spatial radius $r_x$, respectively, are separated by a first order phase transition line. Above a critical temperature a third phase corresponding to a non-uniform eigenvalue distribution arises, while the phase transitions soften to be of second and third order, respectively.

This tremendous progress non-withstanding, the question of topology change unfortunately cannot be answered using simulations of equilibrium quantities, because it requires information about real-time evolution.

Real-time information in quantum field theories are extremely difficult to obtain, because there is no known way to study the real-time dynamics of 
quantum systems at a reasonable computational cost. Absent any breakthrough development in the numerical treatment of real-time quantum systems, we take a more pedestrian approach in the present work by studying the real-time evolution in the \textit{classical statistical} approximation, the non-perturbative lattice technology of which has been well developed in the context of relativistic heavy-ion collisions \cite{Krasnitz:1999wc,Lappi:2003bi,Romatschke:2005pm,Berges:2008zt,Gelfand:2016yho} (see also \cite{Ambjorn:1990pu, Bodeker:1999gx}).

From a gauge theory perspective, the real-time dynamics of fully quantum 2d SYM is expected to be well approximated by its classical dynamics for modes which are highly occupied. For bosons in equilibrium, this is the case for the modes with energy $\omega$ obeying $\beta \omega \ll 1$, or the low-energy modes. At small temporal circle radius $\beta$ or high temperature, many bosonic modes can be well approximated by their classical dynamics. Additionally, at high temperature fermionic modes develop a large thermal mass, so they can be expected to decouple and thus not contribute to observables that are insensitive to the number of degrees of freedom of the theory. Thus at high temperature, one can expect the dynamics of 2d SYM to be at least qualitatively well approximated by the classical dynamics of purely bosonic Yang-Mills theory.
This expectation has indeed be confirmed in previous numerical studies in other dimensions \cite{Asplund:2011qj,Asplund:2012tg,Aoki:2015uha,Kunihiro:2010tg,Gur-Ari:2015rcq} and a comparison of Monte Carlo results between 2d SYM and 2d SU(N) pure gauge theory \cite{Hanada:2016qbz}. (Note that generalizations of the classical statistical framework to include quantum effects to some extent may be possible, cf. \cite{Rebhan:2004ur,Dumitru:2005gp,Arnold:2005vb,Buividovich:2017kfk}.)

Not surprisingly, some aspects of the 2d SYM dynamics cannot be captured by the purely bosonic classical statistical approximation. For instance, by neglecting the fermions one is studying the phase diagram of the purely bosonic theory shown in Fig.~\ref{fig:PhaseDiagram}, cf. Refs. \cite{Aharony:2005ew,Hanada:2007wn,Mandal:2011hb,Hanada:2016qbz}. However, at high temperature, the 2d SYM and pure Yang-Mills phase diagrams become indistinguishable, such that the classical statistical simulations performed as part of this work could reasonably be expected to offer qualitative insights into the black string phase, and potentially also the deconfined black hole phase.

From string theory point of view, the high temperature (weak coupling) regime should describe a highly stringy system, 
where the $\alpha'$ corrections are large but one can still tune the $g_s$ correction by dialing $N$. By performing direct numerical simulations of the high temperature gauge theory dynamics, one can study whether the $\alpha'$ correction is sufficient for resolving the singularity associated with the topology change, and how the $g_s$ correction affects the result. A key asset of our classical statistical simulations is that we should be able to see how a perturbation of a black hole/black string rings down as the system (re-)approaches equilibrium. In the context of classical gravity, this ring-down process is encoded in the quasi-normal mode spectrum \cite{Berti:2009kk}, which we attempt to measure numerically in purely bosonic Yang-Mills. 

This paper is organized as follows: section 
\ref{sec:one} contains background on the simulation method, a particular scaling symmetry and the equilibrium phase diagram. Section \ref{sec:two} discusses rapid quenches of the system energy and corresponding signals of topology change, including quasinormal mode ringing. We summarize and conclude in section \ref{sec:conc}.

\section{Equilibrium Phase Diagram}
\label{sec:one}

We study SU($N$) Yang-Mills theory with 8 scalars on a spatial circle, in Minkowski signature such that the dynamics is 1+1 dimensional. The theory will be set up using the tools from lattice gauge theory, by discretizing 9-dimensional classical Yang-Mills on a lattice where eight dimensions are toroidally compactified (see Ref.~\cite{Hanada:2016qbz} for more discussion about this point).

\subsection{Simulation Method}

In order to solve the equation of motion while preserving gauge invariance it is useful to make use of standard lattice formulations of gauge fields. Our setup is based on our earlier work in Ref. ~\cite{Hanada:2016qbz} on quantum Monte-Carlo simulations, which we briefly review here in order to keep this work self-contained\footnote{The code package ``askja'' that allows lattice simulations of SU($N$) Yang-Mills for arbitrary $N$ in arbitrary number of dimensions with arbitrary toroidal compactification is publicly available at \cite{codedown}. }.

We replace continuum 9-dimensional Euclidean space by an isotropic cubic lattice such that \hbox{$x^i=a \hat{x}^i$}, $i=1,2,\ldots 9$ with $\hat{x}$ taking on integer values and $a$ being the (spatial) lattice spacing. With $g$ denoting the strong coupling constant, the continuum gauge field variables $A_i(x)$ are replaced by link variables \hbox{$U_i(x)=e^{a A_i(x)}=e^{-ig a A_i^a(x) T^a}$} which are elements of the SU($N$) Lie group and live on links between lattice sites $\hat{x}^i$ and obeying $U_i^\dagger(x)=e^{i g a A_i^a(x) T_a}=e^{-a A_i(x)}$.  This allows to use
the standard single plaquette definition for the  Hamiltonian density
\begin{equation}
\label{hdendisc}
{\cal H}=\frac{1}{4}E_i^a E_i^a+\frac{N }{g^2 a^4}\sum_\Box\left(1-\frac{1}{N}{\rm Re\ Tr} U_{\Box,ij}\right)\,.
\end{equation}
where $E_i^a=\frac{d A_i^a(x)}{dt}=\frac{\partial {\cal H}}{\partial E_i^a(x)}$ and  $U_\Box$ is defined through \cite{Bodeker:1999gx,Hanada:2016qbz}:
\begin{equation}
U_{\Box,ij}=U_i(x)U_j(x+i)U_i^\dagger(x+j)U_j^\dagger(x)\,.
\end{equation}
(Note that $\sum_\Box$ denotes the sum over all spatial loops on the lattice starting from site $\hat{x}^i$ with only one orientation, e.g. $
\sum_\Box\equiv \sum_{1\leq i < j \leq d}$.) The lattice equations of motion are calculated from the Hamiltonian equation and one finds
\begin{equation}
\label{eq:Uevol}
U_i(\hat{x},\hat{t}+1)=e^{i \Delta t E_i(x)}U_i(\hat{x},\hat{t})\,,
\end{equation}
when using the definition $E_i(x)=-g a^2 T^a E_i^a(x)$ and discretizing time in units of $t=a \Delta t\, \hat{t}$ with $\hat{t}$ an integer. The 
update rule for the electric field is found by requiring $\dot{H}\equiv \frac{d H(t)}{dt}=0$ \cite{Bodeker:1999gx,Berges:2008zt,Hanada:2016qbz}:
\begin{equation}
\label{eq:Eevol}
E_i(\hat{x},\hat{t}+\frac{1}{2})=E_i(\hat{x},\hat{t}-\frac{1}{2})-\Delta t\sum_{|j|\neq i} {\rm Adj}\left[U_i(\hat{x},\hat{t})S_{ij}^\dagger(\hat{x},\hat{t})\right]\,,
\end{equation}
where the gauge staple $S_{ij}$ is defined as in \cite{Bodeker:1999gx} as 
$S_{ij}(x)=U_j(x) U_i(x+j) U_j^\dagger(x+i)$ and ${\rm Adj}\left[M\right]\equiv - \frac{i}{2}\left[M-M^\dagger-\frac{1}{N}{\rm Tr}\left(M-M^\dagger\right)\right]$ for SU($N$).
Note that for negative values of $j$, a gauge link is traversed in the opposite direction, e.g. $U_{-j}(x)=U_j^\dagger(x-j)$. The Hamiltonian for the system is given by
\begin{equation}
\label{eq:Etot}
H(\hat{t})=\frac{N a^{d-4}}{g^2}\sum_x\left[\frac{{\rm Tr}\left[\left(E_i(\hat{x},\hat{t}+\frac{1}{2})+E_i(\hat{x},\hat{t}-\frac{1}{2})\right)^2\right]}{8 N}+\sum_\Box \left(1-\frac{1}{N}{\rm Re\ Tr} U_{\Box,ij}(\hat{t})\right)\right]\,,
\end{equation}
where $E_i^2\equiv E_i E_i^\dagger$. Similarly, the Gauss law constraint on the lattice is given by
\begin{equation}
\label{eq:gausslaw}
G(\hat{t})=\sum_i \left(E_i(\hat{x},\hat{t}+\frac{1}{2})-U_i^\dagger(\hat{x}-\hat{e}_i,\hat{t}) E_i(\hat{x}-\hat{e}_i,\hat{t}+\frac{1}{2})U_i(\hat{x}-\hat{e}_i,\hat{t})\right)\simeq 0\,.
\end{equation}

To prepare initial conditions which satisfy $G(\hat{t})=0$, we simply take $E_i=0$ and start with link variables $U_i(\hat x)=\exp(-i g a A_i^a(\hat x) T^a)$, where $A_i^a(\hat{x})$ are Gaussian-random with predefined magnitude.  Note that this is different than the initial conditions chosen for standard (quantum, Euclidean) lattice Monte Carlo simulations. Once initial conditions are specified, we determine the total system energy from measuring (\ref{eq:Etot}) and real time evolution on the lattice is then performed by using set of equations (\ref{eq:Uevol}), (\ref{eq:Eevol}) to time-step forward the fields $U_i,E_i$. Our evolution scheme is accurate up to ${\cal O}(a^2)$ corrections in time.

One of the main observables in this work will be the Wilson loop defined as a product over link matrices
\begin{equation}
  \label{eq:wdef}
  {\cal W}\equiv\prod_{m=1}^{m=r_x a} U_x\left(m\right)\,.
\end{equation}
We will study the absolute value of the normalized trace of the Wilson loop,
\begin{equation}
  |W|\equiv \frac{1}{N}{\rm Tr}\, {\cal W}\,,
\end{equation}
as well as the distribution $\rho(\theta)$ of the Eigenvalue spectrum $e^{i \theta_1},e^{i \theta_2},\ldots, e^{i \theta_N}$ of ${\cal W}$ with $-\pi\leq \theta\leq \pi$.

It should be pointed out that for spatial directions with only one site and periodic boundary conditions, the corresponding gauge field becomes a scalar, e.g. $A_i\rightarrow X_i$. In the following we consider the situation where eight spatial directions are compactified on a point, while the ninth direction is allowed to be large, so that $A_i\rightarrow \left\{X_I,A_x\right\}$ with $I=1,2,\ldots 8$.

\subsection{Relation to Matrix Models and Scaling Symmetry}
\label{sec:symm}

In continuum, the lattice-discretized theory described above corresponds to the classical Lagrangian
\begin{eqnarray}
L
=
\frac{1}{g^2}\int_0^{r_x} dx {\rm Tr}
\left\{
-\frac{1}{4}F_{\mu\nu}F^{\mu\nu}
-\frac{1}{2}(D_{\mu}X_I)(D^{\mu}X_I)
+\frac{1}{4}[X_I,X_J]^2
\right\}, 
\end{eqnarray}
where $\mu=\left\{t,x\right\}$ and $F_{\mu\nu},D_\mu$ are the two-dimensional field strength and gauge covariant derivative, respectively. In the 't Hooft large-$N$ limit, the 't Hooft coupling $\lambda=g^2N$, the compactification radius $r_x$ and the energy  per d.o.f $E/N^2$
are fixed to be of order $N^0$. Since in two dimension $\lambda$ is dimensionful, in the following we employ dimensionless units such as
\begin{equation}
  x\rightarrow \lambda^{1/2} x\,, \quad
  t\rightarrow \lambda^{1/2} t\,, \quad
  X_I\rightarrow \lambda^{-1/2} X_I\,,\quad
  L\rightarrow \lambda^{1/2} L\,,\quad
  E\rightarrow \lambda^{-1/2} E\,.
\end{equation}

In temporal gauge $A_t=0$, the classical equations of motion in continuum become
\begin{eqnarray}
\frac{d^2 X_I}{dt^2}
&=&
D_x^2X_I
+
[X_J,[X_I,X_J]], 
\nonumber\\
\frac{d^2 A_x}{dt^2}
&=&
i[D_xX_J,X_J]\,,
\label{sec:2d-EOM} 
\end{eqnarray}
which obey the following scaling symmetry: If $A_x(t,x), X_I(t,x)$ is a solution, then $A'_x(t,x)=\alpha A_x(\alpha t, \alpha x)$, 
$X'_I(t,x)=\alpha X_I(\alpha t,\alpha x)$ with a compactification period $r'_x=\alpha^{-1} r_x$ is also a solution, 
with the energy $E'=\alpha^3 E$. 
With this rescaling, we can set $r_x=1$. 
Therefore, the energy $E$ and the number of colors $N$ are the only relevant simulation parameters.


\subsection{UV Problem in Classical Statistical Lattice Simulations}

Classical statistical simulations suffer from a well-known ultraviolet instability (this is the same as the Rayleigh-Jeans law for black body radiation which is cured by quantum mechanics). Given a finite lattice discretization scale $a$ (the lattice spacing), classical dynamics will eventually start to populate modes close to the lattice UV cutoff scale $a^{-1}$ at late times. The dynamics of these high momentum modes, however, should involve quantum effects which are absent in the classical statistical simulations. Therefore, once a simulation starts to become sensitive to the lattice UV scale, the resulting dynamics can no longer be trusted, and the simulation has to be stopped.

\begin{figure}[t]
  \begin{center}
    \includegraphics[width=.49\linewidth]{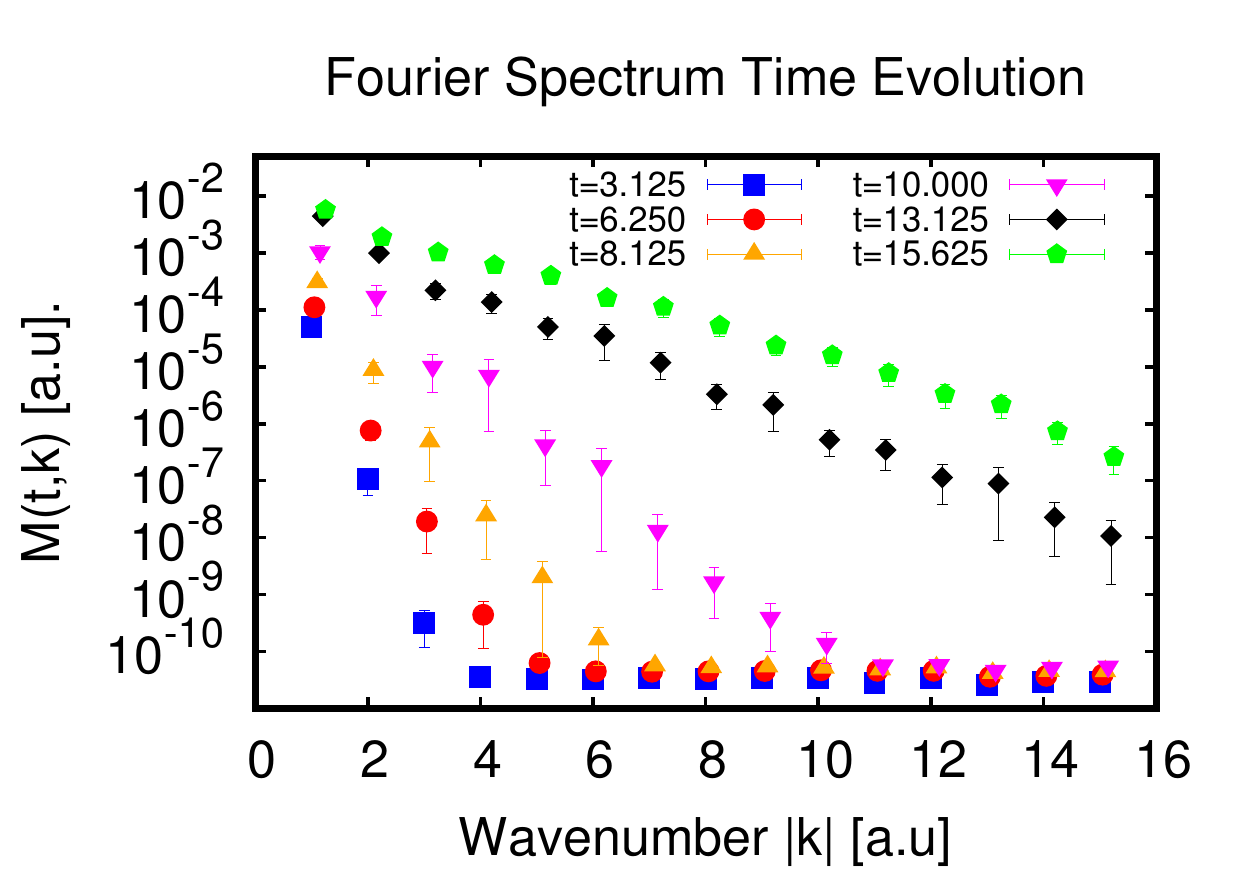}
    \includegraphics[width=.49\linewidth]{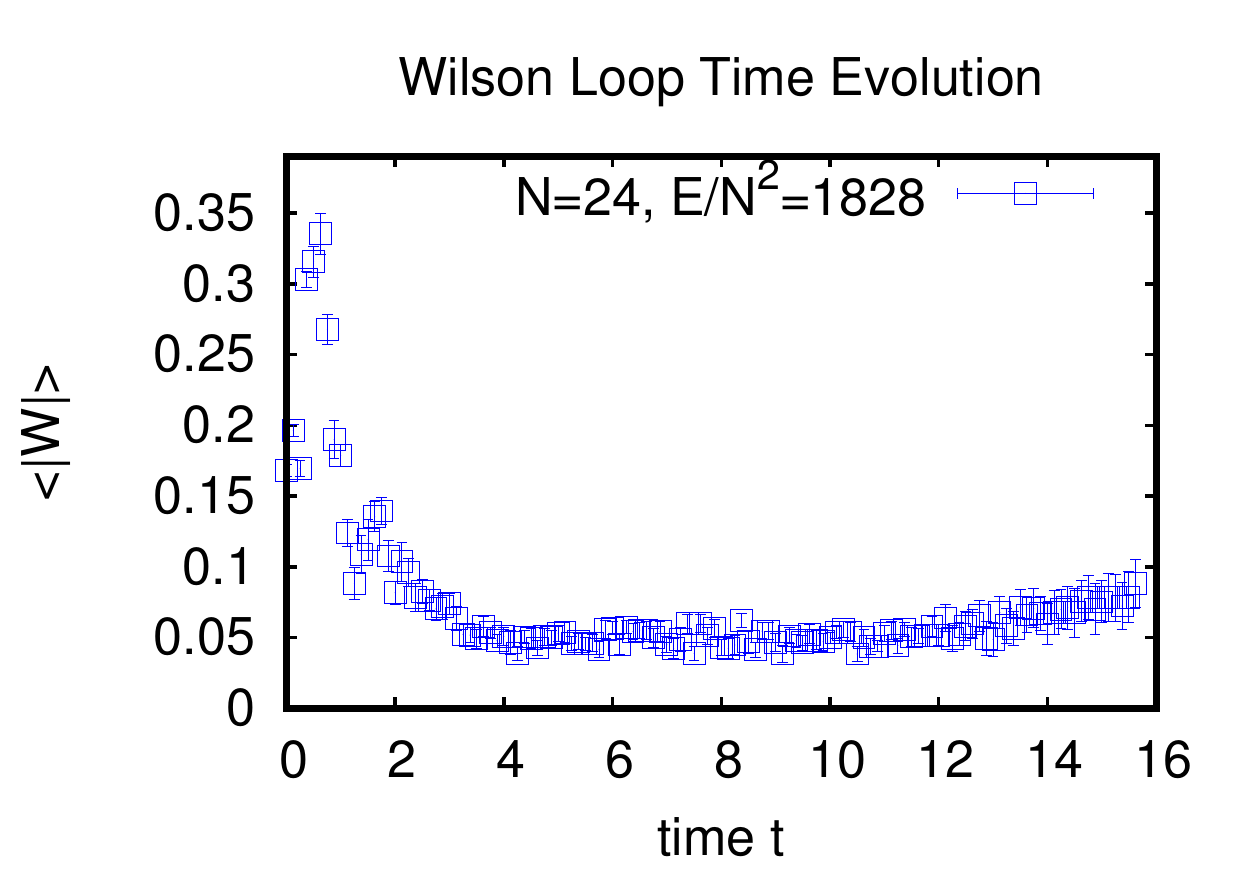}
            
  \end{center}
  \caption{%
    Left: Time-evolution of magnitude $M(\hat t,\hat k)$ of Fourier modes as a function of momentum. At times $t\leq 10$, $M(\hat t,\hat k)$ only has support in the infrared, while for $t \geq 13$, pile-up of modes in the UV occurs.  Right: Time-evolution of absolute Wilson loop expectation value $\langle |W|\rangle$. After some initial transient for $t\leq 4$, $\langle |W|\rangle$ is approximately constant over time for times $4 \leq t\leq 12$. Based on the information from the Fourier modes shown in the left plot, the behavior of $\langle |W|\rangle$ for $t\geq 12$ seems to be contaminated by lattice artefacts resulting from mode pile-up in the UV. 
    All simulations for a 32 site lattice with $N=24$, $\Delta t=0.01$, and a normalized energy $E/N^2\simeq 1828$; error bars shown represent statistical error over ensembles.
  }\label{fig:Fourier-N48}
\end{figure}

Nevertheless, classical lattice statistical simulations with a fixed UV cutoff $a^{-1}$ can successfully be used for system properties dominated by modes in the IR. In practice, we prepare initial conditions for the simulations which are well localized in the IR, and then run the simulations for sufficiently short times before modes start to pile-up at the UV lattice cutoff scale. To check if pile-up has occurred, we monitor the Fourier transform of the magnetic component of Hamiltonian,
\begin{equation}
M(\hat t, \hat k)\equiv  \sum_{\hat x} e^{i \hat k \hat x} \sum_\Box \left(1-\frac{1}{N}{\rm Re\ Tr} U_{\Box,ij}(\hat{t})\right)\,,
\end{equation}
which has successfully been used as an indicator for this purpose in the past \cite{Romatschke:2005pm,Romatschke:2006nk}. A representative plot of the time-evolution of $M(\hat t,\hat k)$ is shown in Fig.~\ref{fig:Fourier-N48}, from which it can be seen that the system starts out dominated by IR physics, but eventually flows to the UV. As long as the system is dominated by IR physics, observables such as the expectation value of the Wilson loop (\ref{eq:wdef})
are unaffected by the classical UV instability, but become sensitive as soon as pile-up of modes in the UV occurs, as shown in the right panel of Fig.~\ref{fig:Fourier-N48}. For this reason, the results reported below have been obtained by only using simulation data for times $t<t_{\rm UV}$, where $t_{\rm UV}$ denotes the time when UV pile-up has first occurred.

\subsection{Results for Equilibrium Phase Diagram in Classical Statistical Simulations}

\begin{figure}[htbp]
  \begin{center}
     \rotatebox{-90}{
   \includegraphics[width=70mm]{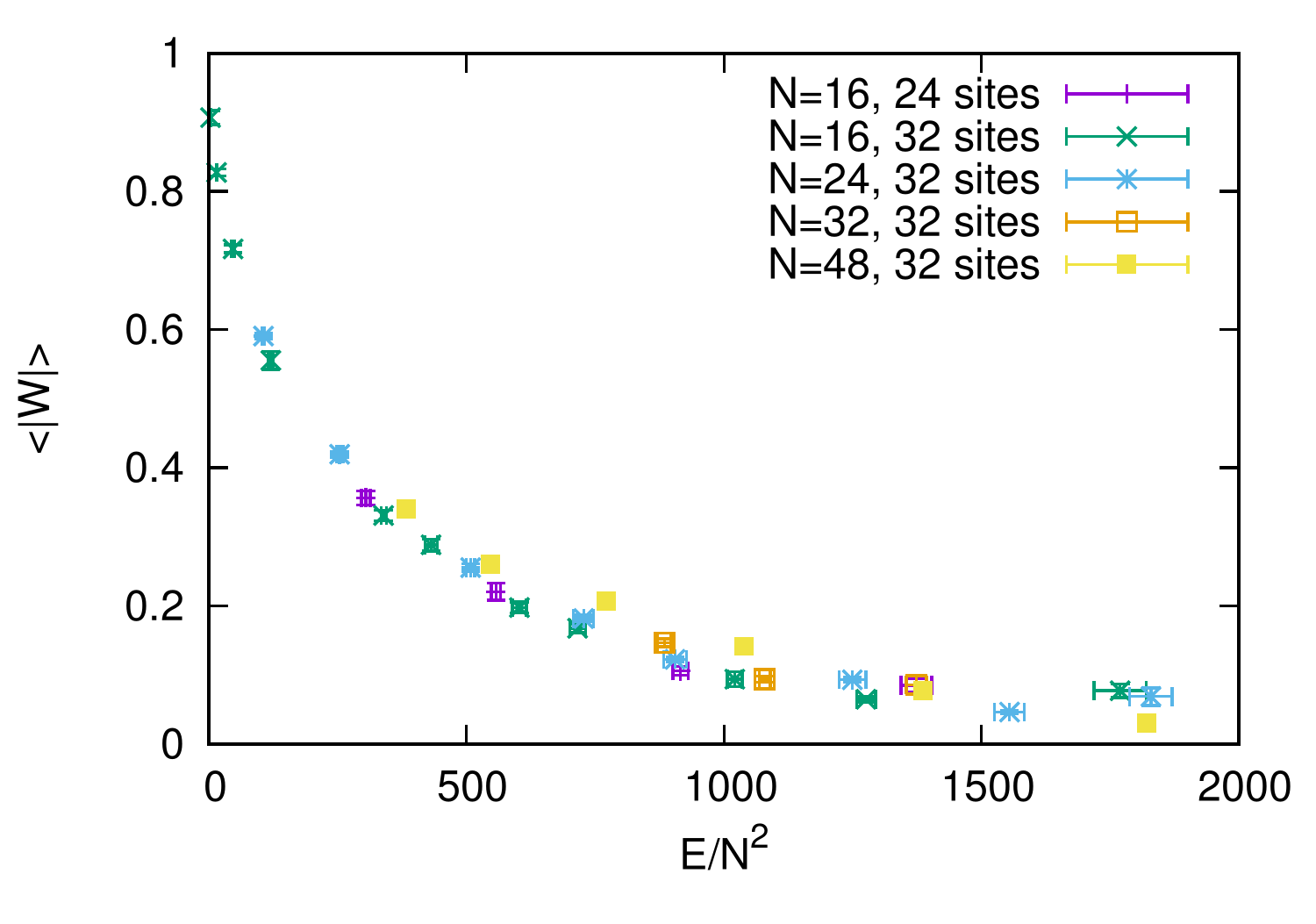}}
  \end{center}
  \caption{Equilibrium Wilson loop expectation value versus energy in classical statistical simulations for various values of $N$ and two lattice sizes.  }\label{fig:W-vs-E}
\end{figure}

As a reminder, classical statistical simulations are performed in the microcanonical ensemble (at fixed energy $E$) instead of the grand-canonical ensemble (fixed temperature $T$) employed in quantum simulations of lattice field theory. Real-time information on observables may be obtained by averaging time-dependent results over classical ensembles. In order to connect to equilibrium properties of the system, one would want to measure observables after the system has become time-independent (thermalized) at time $t\geq t_{\rm therm}$, but before the simulation becomes contaminated by UV artefacts $t\leq t_{\rm UV}$. We find that both $t_{\rm therm}$ and $t_{\rm UV}$ depend on the normalized energy $E/N^2$ in simulations.

In practice, for fixed $E/N^2$ we consider the time-evolution of observables such as the ensemble-averaged Wilson loop expectation value shown in Fig.~\ref{fig:Fourier-N48}, and perform an additional time-average over $t_{\rm therm}\leq t\leq t_{\rm UV}$ to increase statistics. Results from this procedure for the Wilson loop expectation value as a function of energy for various $N$ are shown in Fig.~\ref{fig:W-vs-E}. We find that results for different $N\geq 16$ cluster around an apparently universal curve, indicating that $\langle |W|\rangle(E/N^2)$ is approximately independent of $N$.

At large $E/N^2$, where our classical-statistical simulations can reasonably be expected to be a good approximation of the full quantum results, we find that $\langle |W|\rangle\rightarrow 0$, in accordance with bosonic quantum theory simulations \cite{Hanada:2016qbz}. Conversely, bosonic quantum theory simulations indicate that $\langle |W|\rangle(E/N^2)\neq 1$ for low temperature, whereas from Fig.~\ref{fig:W-vs-E} it can be seen that classical statistical simulations result in $\langle |W|\rangle(E/N^2)\rightarrow 1$ for $E\rightarrow 0$, because all link variables are equal to the identity matrix in this limit. Furthermore, full quantum theory simulations exhibit confinement for low temperatures  \cite{Hanada:2016qbz}, whereas classical statistical simulations do not show confinement. Therefore, as expected, classical statistical simulations can not be used to study properties of, or the transition to, the confined center broken phase of the full bosonic quantum theory (cf. Fig.~\ref{fig:PhaseDiagram}).

\begin{figure}[t]
  \begin{center}
    \includegraphics[width=.9\linewidth]{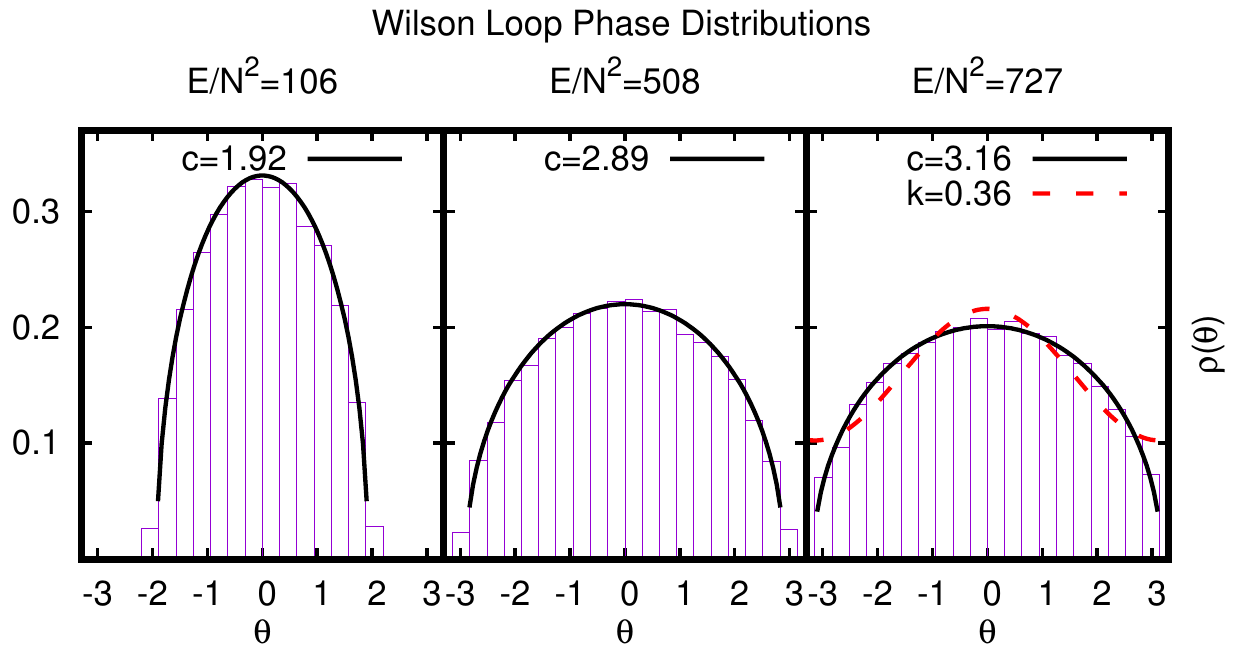}
    \includegraphics[width=.9\linewidth]{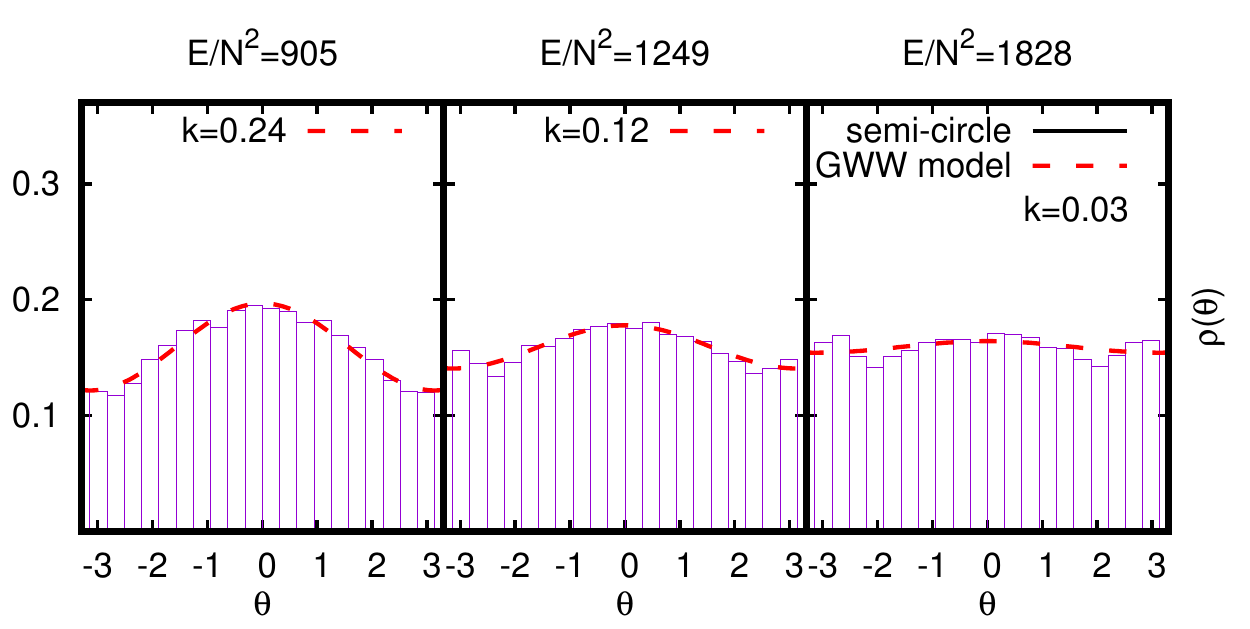}
  \end{center}
  \caption{Wilson loop phase distribution $\rho(\theta)$ for various energies. For $E/N^2\leq 750$, the results can be fit well by a semi-circle distribution (\ref{eq:semici}) with fit parameter $c$, whereas for $E/N^2\simeq 900$, results are well fit in terms of the Gross-Witten-Wadia (GWW) model (\ref{eg:GWWm}) with fit parameter k. Best-fit values for $c,k$ are indicated in individual plots panels. }\label{fig:Wilson-phases}
\end{figure}

However, one may ask if classical statistical simulations can be used to probe the properties of the quantum theory in the deconfined phase, notably the region close to the center symmetry phase transition shown in Fig.~\ref{fig:PhaseDiagram}. In order to study the phase structure more precisely,  the distribution of the Wilson line phases $\rho(\theta)$ are shown in Fig.~\ref{fig:Wilson-phases} as a function of energy.

At high energy, one finds that the phase distribution $\rho(\theta)$ in the classical statistical simulations is ungapped, see Fig.~\ref{fig:Wilson-phases}. This is consistent with the full bosonic quantum simulations \cite{Hanada:2016qbz}, and as such is consistent with the conjectured gravity picture of a black string phase at high temperature that was eluded to in the introduction.

Conversely, at low energy $E/N^2\rightarrow 0$, we see clear evidence for a gapped distribution of Wilson loop phases in the classical statistical simulations that can be well-fit by a semi-circle distribution
\begin{equation}
  \label{eq:semici}
  \rho_{\rm semi-circle}(\theta)=\frac{\sqrt{c^2-\theta^2}}{\pi c^2/2}\,,
\end{equation}
with one fit parameter $c$ with best-fit values  indicated in Fig.~\ref{fig:Wilson-phases}.

\begin{figure}[t]
  \begin{center}      
   \includegraphics[width=.7\linewidth]{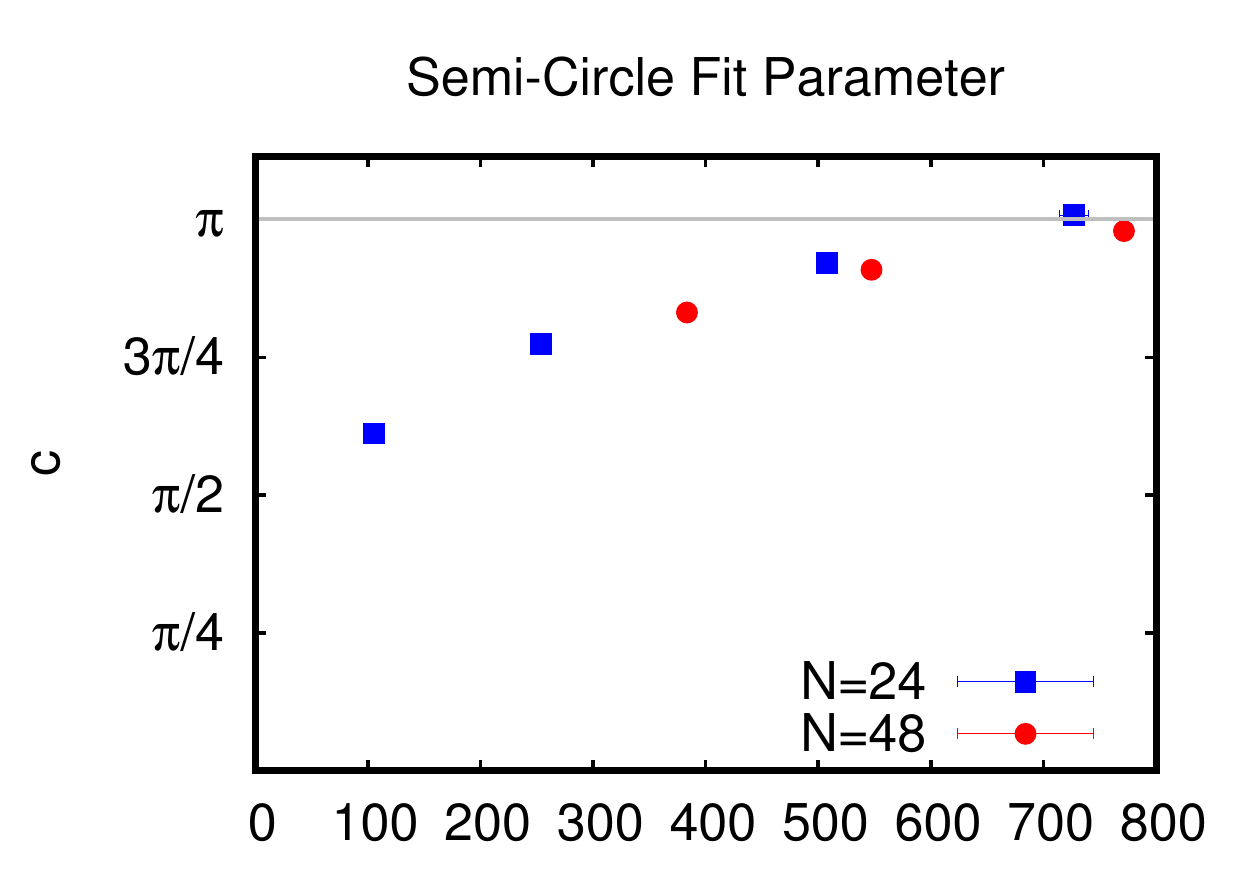}
  \end{center}
  \caption{The coefficient $c$ obtained by fitting the classical-statistical Wilson loop phase distributions for various energies by the semi-circle distribution Eq.~(\ref{eq:semici}) for $N=24$ and $N=48$ on 32 site lattice. Grey line is a guide to the eye. Results seem to suggest that $c\rightarrow \pi$ around $E/N^2\simeq 750\pm50$ independent from $N$. }\label{fig:c-vs-E}
\end{figure}

As indicated in Fig.~\ref{fig:PhaseDiagram}, a phase transition from ungapped to gapped Wilson line phase is expected as the temperature is lowered at fixed circle length in the full bosonic quantum theory. (In SYM, the transition in the microcanonical ensemble has been found to be first order in Ref.~\cite{Dias:2017uyv}). A priori, it is not obvious that such a transition should be accessible using classical statistical simulations. However, the fact that the behavior of the Wilson loop eigenvalues changes qualitatively between low and high energy suggests that classical statistical simulations are able to access both the ungapped and gapped deconfined phase separated by either a phase transition or rapid analytic crossover transition, respectively. Thus, while classical statistical simulations cannot be expected to be quantitatively accurate approximations in the gapped phase, our results strongly suggest that they can be used to gain results which are qualitatively similar to the full quantum theory, including the real-time dynamics close to the phase boundary of the ungapped phase.

The gapped Wilson loop phase distribution observed at low energy in the classical statistical simulation is consistent with the conjectured gravity dual picture of a localized black hole configuration. As the energy in the classical statistical simulations is increased, the fit of $\rho(\theta)$ in terms of the semi-circle distribution (\ref{eq:semici}) leads to increasing fit parameter $c$ (see Fig.~\ref{fig:c-vs-E}), where $c=c_{\rm crit}=\pi$ having the natural interpretation\footnote{The fact that the semi-circle fit shown in Fig~\ref{fig:Wilson-phases} works so well even close to the presumptive phase transition at $c=\pi$ could be an indication that the black hole phase on $S^1$ is well approximated by the black hole solution in a non-compact space, cf. Ref.~\cite{Kudoh:2004hs}.} of separating black-hole from black string configurations at a critical normalized energy of
\begin{equation}
  \label{eq:critE}
\left(E/N^2\right)_{\rm critical}\simeq 750\pm 50\,,
\end{equation}
where the uncertainty is primarily coming from the residual $N$-dependence of $c$, cf. Fig~\ref{fig:c-vs-E}. Note that at for $c_{\rm crit}=\pi$, the expectation value for the Wilson loop becomes
\begin{equation}
  \label{eq:critW}
  \langle |W|\rangle_{\rm crit}=\int \rho_{\rm semi-circle}(\theta)e^{i \theta}=\frac{2 J_1(\pi)}{\pi}\simeq 0.181\,.
\end{equation}
We find that for $E/N^2\geq \left(E/N^2\right)_{\rm critical}$, the Wilson loop distribution is rather well fit by another one-parameter form given by 
\begin{equation}
  \label{eg:GWWm}
  \rho_{GWW}(\theta)=\frac{1+ k\cos\theta}{2 \pi}\,,
\end{equation}
which is reminiscent of the analytically known distribution for the non-uniform string \cite{Kawahara:2007fn,Azeyanagi:2009zf} in the Gross-Witten-Wadia model 
\cite{Gross:1980he,Wadia:1980cp}. Best-fit values for $k$ are indicated in Fig.~\ref{fig:Wilson-phases}.

Our current results are not precise enough to say anything about the nature of the transition from gapped to ungapped Wilson loop phase in classical statistical simulations, nor can we confirm or rule out the presence of a third phase at high temperature expected from gravity, namely uniform Wilson loop phase distributions (cf. Fig.~\ref{fig:BH-BS}).

\section{Real-Time Response to Quenches}
\label{sec:two}

The results from the previous section strongly suggests that classical statistical simulations may be used to probe the properties of the black-hole/black-string transition that are in qualitative, if not quantitative, agreement with full quantum theory simulations in equilibrium.

Unlike current quantum simulations, no obstacle prevents the application of classical statistical simulations to non-equilibrium problems, suggesting that our method can be used to probe the real-time dynamics of the topology change from black hole to black string phases. Assuming that the real-time evolution of the Wilson loop phase distribution $\rho(\theta,t)$ possesses a good large-$N$ limit, then such topology changes may take place within a finite time even at large-$N$.

\afterpage{\clearpage}
\begin{figure}[p]
  \begin{center}
    \includegraphics[width=.8\linewidth]{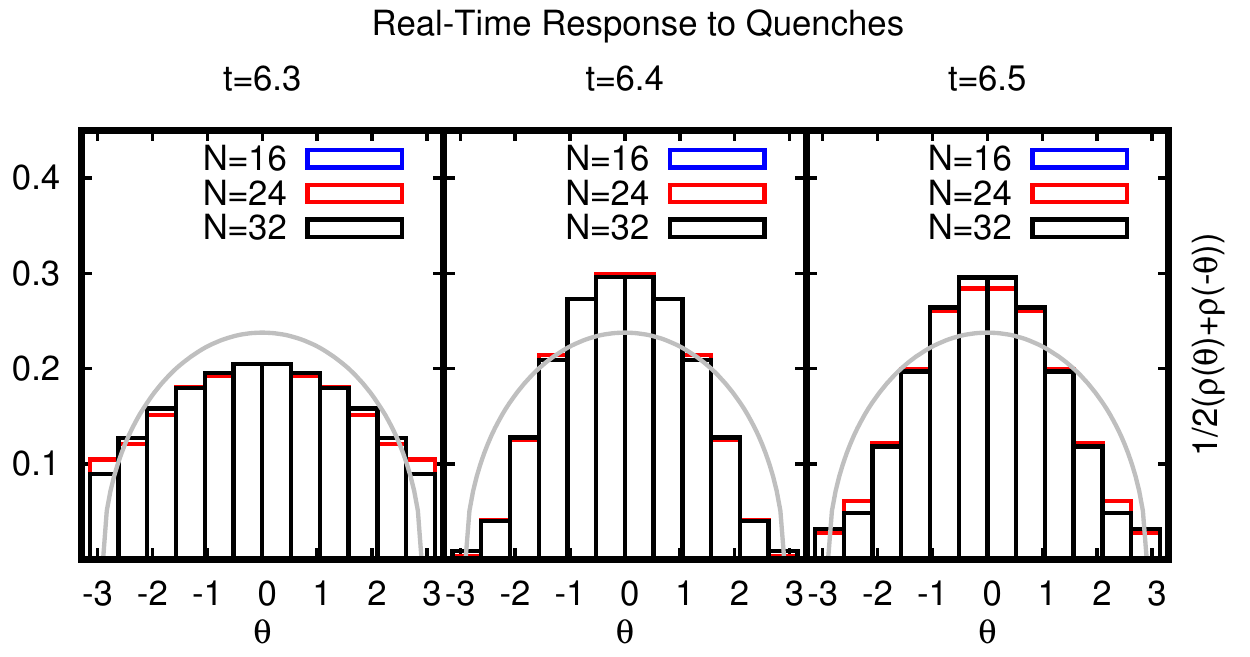}
    \includegraphics[width=.8\linewidth]{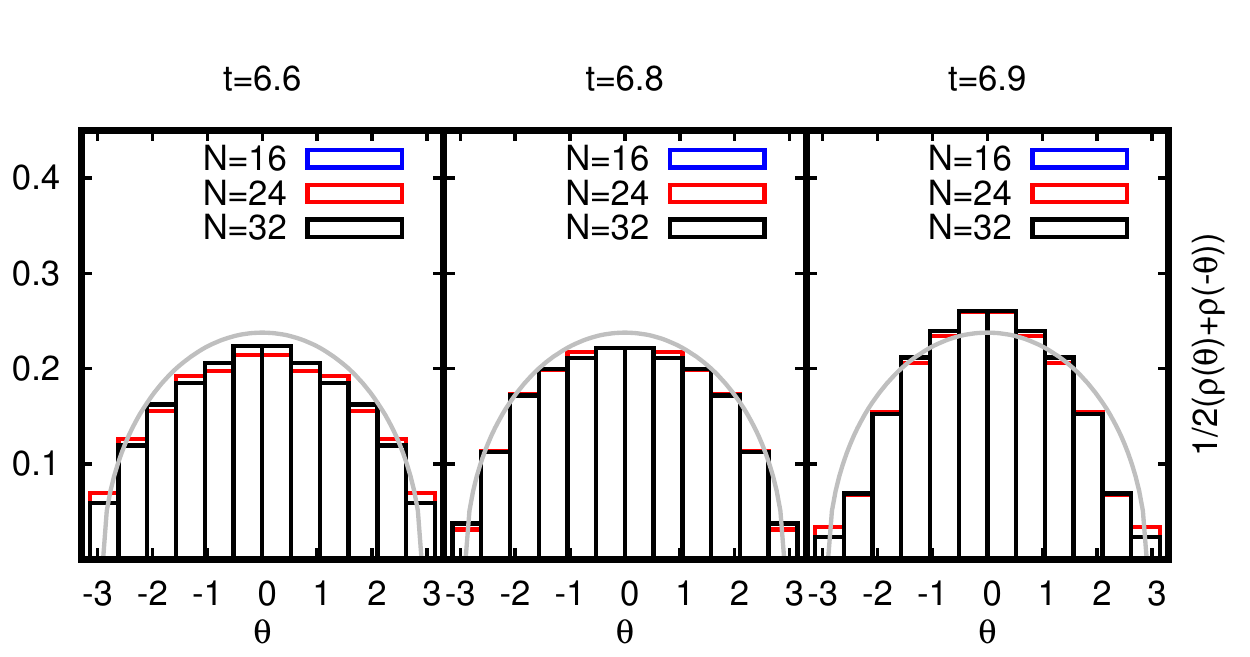}
    \includegraphics[width=.8\linewidth]{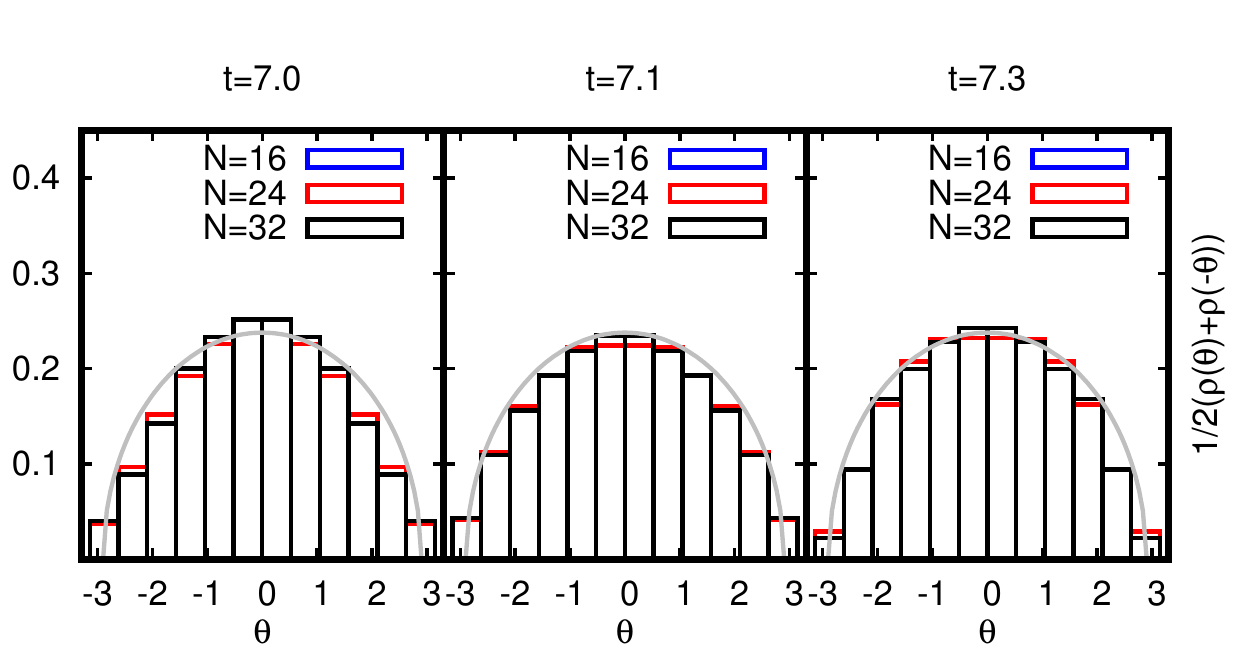}
  \end{center}
  \caption{Real-time response (left to right, top to bottom) of Wilson loop phase distributions to quench from $E/N^2=1200$ (black-string phase) to $E/N^2=500$ (black hole-phase) for $N=16,24,32$, suggesting that results are almost independent from $N$.  Quench was performed at $t=6.25$ and grey line indicates new expected equilibrium configuration (\ref{eq:semici}) with $c=2.92$ based on $E/N^2=500$. }\label{fig:Wilson-rt}
\end{figure}

\subsection{Black-Hole/Black-String Topology Change}

In order to study topology changes in a controlled manner in classical statistical simulations, we employ the following protocol:
\begin{enumerate}
\item
  Generate a classical statistical gauge field configuration with initial energy $E/N^2<750$  ($E/N^2>750$) expected to be in the black hole (black string) phase
\item
  Evolve the gauge field configuration until $t\geq t_{\rm therm}$ so that early-time transients have disappeared
\item
  Rapidly quench the system energy to a fixed final value of $E/N^2>750$ ($E/N^2<750$) without changing the Wilson line phases distribution or violating the Gauss law constraint \eqref{eq:gausslaw}. Note that the new energy indicates an equilibrium configuration in the respective other phase. This quench can be achieved by multiplying the electric field by a constant $q$,
  $$
  E_i(\hat{x},\hat{t})\to q E_i(\hat{x},\hat{t})\,.
  $$
\item
  Measure the real-time response of observables as the system tries to attain the new equilibrium black string (black hole) configuration
\end{enumerate}
Repeating the above protocol for many configuration with fixed initial energy and averaging over this classical ensemble of configurations leads to real-time results for observables shown in the following.

Representative plots for the real-time evolution of the Wilson loop phase distribution are shown in Fig.~\ref{fig:Wilson-rt}, with the corresponding Wilson loop expectation value shown in Fig.~\ref{fig:Wilsosfn-rt}. For these figures, initial configurations with $E/N^2=1200$ were prepared in the black-string phase for different values of $N=16,24,32$. At time $t=6.25$, the system energy was rapidly changed to $E/N^2=500$ and snapshots of the real-time evolution of the phase distribution for times after the quench are shown in Fig.~\ref{fig:Wilson-rt}.  We find that real-time results for the phase distributions show little sensitivity to the choice of $N\geq 16$. Figs.~\ref{fig:Wilson-rt}, \ref{fig:Wilsosfn-rt}  suggest that the Wilson loop phase distribution initially exhibits large-scale fluctuations around the new expected black-hole equilibrium configuration (indicated by a grey semi-circle in Fig.~\ref{fig:Wilson-rt}). These fluctuations decrease in amplitude with time until at times $t\gtrsim 7.3$ the Wilson loop phase distribution is well-described by the semi-circle distribution with $c=2.92$ expected for an equilibrium configuration at $E/N^2=500$.

We have repeated the above procedure to study the real-time response of other quenches, verifying in particular that it is also possible to observe the inverse
process of gapped phase to ungapped phase, which we associate with a topology change process from black-hole to black-string configurations. 

\subsection{Possible Observation of Quasinormal Modes}

\begin{figure}[t]
  \begin{center}
    \includegraphics[width=.8\linewidth]{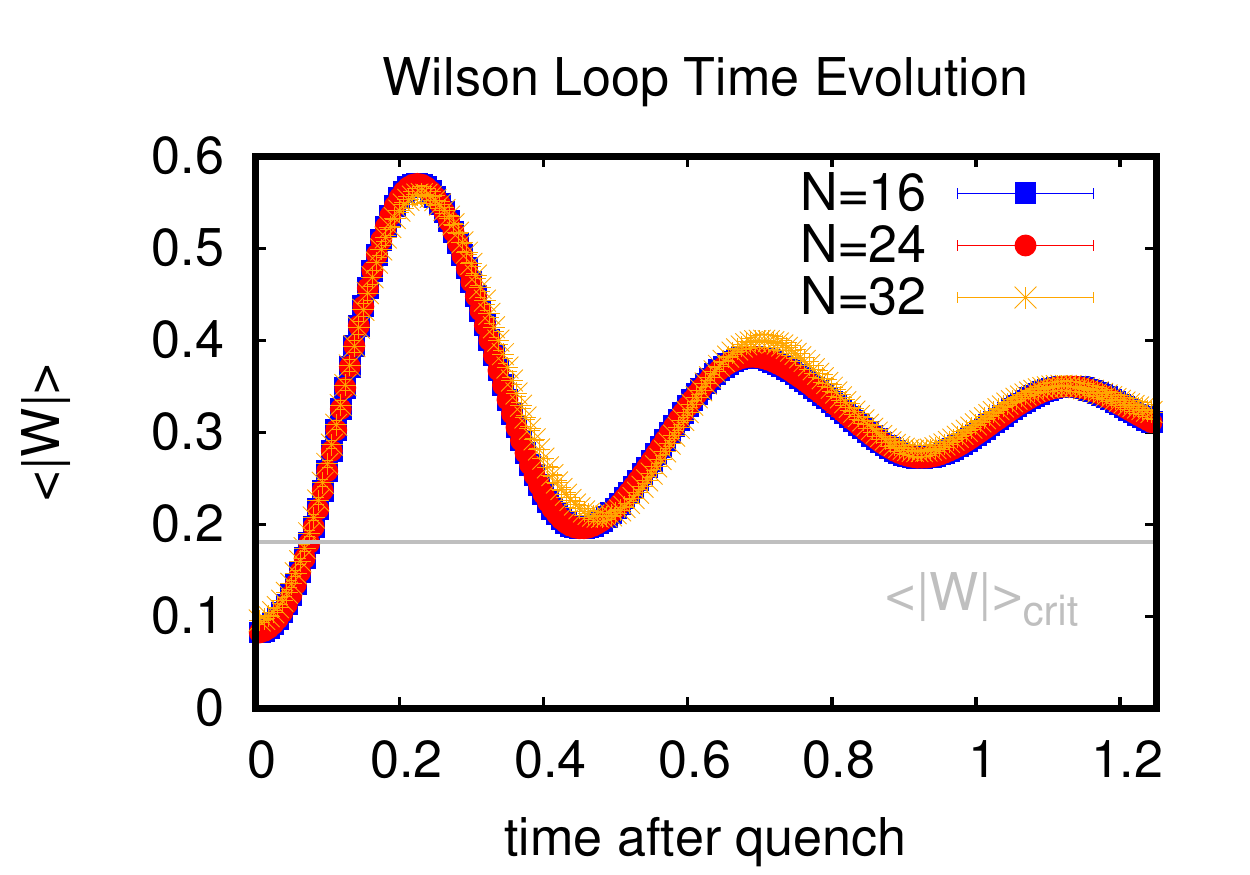}
  \end{center}
  \caption{Time evolution of Wilson loop expectation value after a quench from $E/N^2=1200$ to $E/N^2=500$ for various $N$. Grey line indicates critical value of Wilson loop, Eq.~(\ref{eq:critW}).}\label{fig:Wilsosfn-rt}
\end{figure}

The real-time evolution of Wilson loop phase distribution ensemble-averages show oscillations around the new equilibrium configuration after a rapid quench, cf. Fig.~\ref{fig:Wilson-rt}. These oscillations are particularly visible in the ensemble-averaged Wilson loop expectation value shown in Fig.~\ref{fig:Wilsosfn-rt}, and are apparently displaying little sensitivity to the choice of $N\geq 16$.
Similar oscillations in classical statistical simulations are ubiquitous, with frequency and damping rates associated with the mass and width of quasi-particles that are in good agreement with perturbative quantum field theory, cf. Refs.~\cite{Romatschke:2005pm,Boguslavski:2018beu}.

By contrast, in classical gravity, oscillations of excited geometries for instance of black holes are characterized in terms of their quasinormal mode ringdown behavior, cf. Ref.~\cite{Berti:2009kk}. If the conjectured relation between Wilson loop phase distribution and geometry holds, this naturally leads to the interpretation of linking  quasiparticle oscillations in gauge theory with quasinormal mode oscillations of black holes in string theory.

\begin{figure}[t]
  \begin{center}
    \includegraphics[width=.49\linewidth]{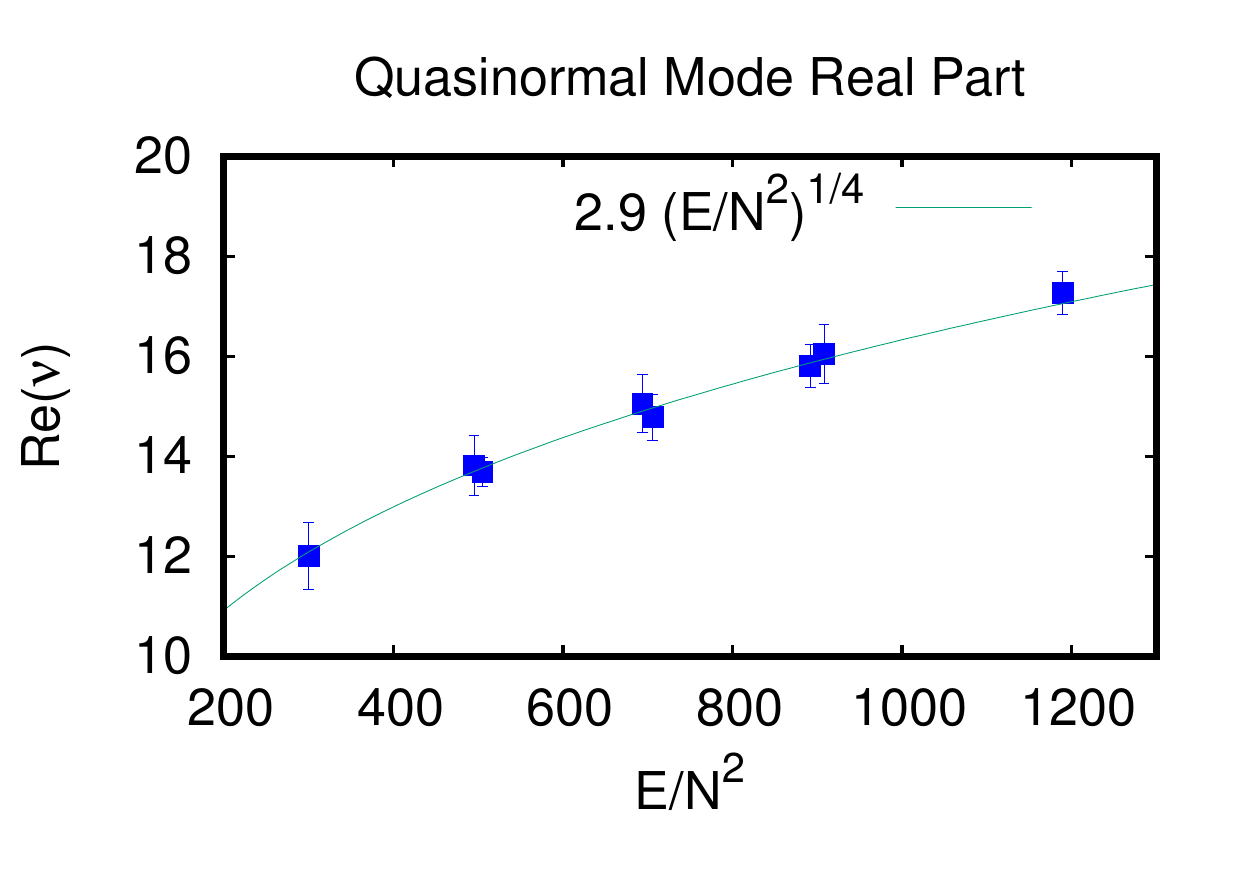}
    \includegraphics[width=.49\linewidth]{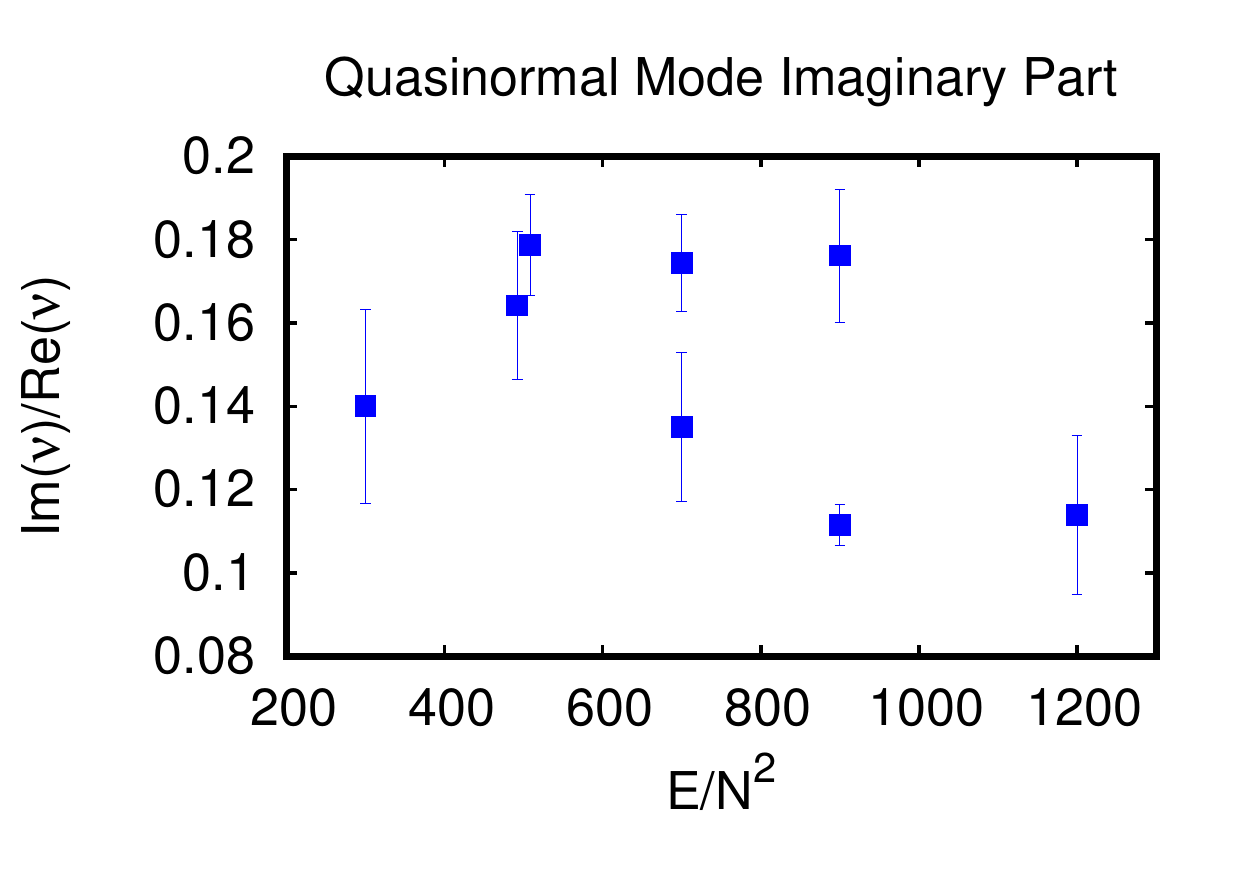}
  \end{center}
  \caption{Real and imaginary part of lowest-lying  presumptive quasinormal mode frequency $\nu$ as a function of final system energy for $N=32$. Multiple entries correspond to quenching protocols with different initial, but the same final energy, e.g. $E/N^2=500\rightarrow 700$ and $E/N^2=900\rightarrow 700$, and have been slightly displaced on the plot to increase visibility.}\label{fig:quasis}
\end{figure}

We are able to obtain estimates for both the real and imaginary part of the lowest-lying  mode frequency $\nu$ by fitting the location and height of the extrema of $\langle |W|(t)\rangle$ to a form $\langle |W|(t)\rangle\propto {\rm Re} \left(e^{i \nu t}\right)$. Results from this fitting procedure for various final energies $E/N^2$ are shown in Fig.~\ref{fig:quasis}. Also shown in in Fig.~\ref{fig:quasis} are results from the fitting procedure when changing the initial energy but leaving the final energy after the quenched unchanged. Our results seem to imply that the extracted results for $\nu$ show little sensitivity to initial system energies, instead only depending on the final $E/N^2$. The behavior of $\nu$ seems to be smooth and continuous with energy, and as such is apparently insensitive to the phase change near (\ref{eq:critE}). Furthermore, our results for $\nu$ are consistent with those from matrix model simulations which do not suffer from UV problems \cite{Aprile:2016mis}, to which our simulations reduce to in the zero volume limit.
The smooth analytic behavior of $\nu$ on $E/N^2$ in classical statistical simulations may be obtained from the classical scaling symmetry outlined in section \ref{sec:symm}. Under the scaling symmetry, frequencies are expected to scale as $\nu\rightarrow \alpha \nu$, and the energy density scales as $\epsilon\rightarrow \alpha^4 \epsilon$. At fixed temperature, assuming the energy density in classical statistical simulations to be approximately independent of volume, we therefore expect $\nu\propto \left(\frac{E}{N^2}\right)^{1/4}$ at fixed volume.
Note that this scaling argument would apply to any time-dependent quantity, such as 
for example for the Lyapunov exponent $\lambda_L\propto (E/N^2)^{1/4}$, cf. Ref.~\cite{Gur-Ari:2015rcq}.

Results shown in Fig.~\ref{fig:quasis} indicate that $\nu$ increases with energy, qualitatively consistent with the finding that the quasinormal mode frequency calculated in type II supergravity on black $p$-brane background, increases with temperature regardless of the value of $p$ \cite{Iizuka:2003ad}. Quantitatively, the quasinormal mode frequency in type II supergravity for $p=1$ differs from the the classical statistical results shown in Fig.~\ref{fig:quasis}, which is expected because supergravity results are applicable for small temperatures whereas classical statistical approximations are quantitatively accurate at high temperatures.



\section{Summary and Conclusions}
\label{sec:conc}

In the present work, we have performed classical statistical simulations of microcanonical ensembles in SU($N$) Yang-Mills theory with eight scalars on a circle. Depending on the energy, we have identified two distinct equilibrium phases of the Wilson loop eigenvalue distribution which qualitatively correspond to those expected for black holes and black strings in the conjectured dual gravity picture. We found that gapped Wilson loop equilibrium phase distributions occurred for energies $E<E_{\rm crit}$, while ungapped distributions occurred for $E>E_{\rm crit}$, with our estimate for $E_{\rm crit}$ given in (\ref{eq:critE}). Our present results were not precise enough to decide if there is an actual phase transition at $E=E_{\rm crit}$ as opposed to an analytic cross-over in classical statistical simulations, nor can we confirm or rule out the presence of second transition to a phase of uniform Wilson loop distributions at very high energies.

We were able to perform real-time measurements of Wilson loop distributions following a quench in system energy from one phase to another. We found characteristic oscillations in the phase distribution and the Wilson loop expectation value, which showed very little sensitivity on the number of colors for $N\geq 16$ or the state the system was in before the quench. Interpreting these oscillations as the string-theory analogue of quasinormal mode ringdown of black holes in classical gravity, we were able to extract estimates for the real and imaginary part of the lowest-lying presumptive quasinormal mode as a function of energy. Our results were found to be in qualitative, but not quantitative, agreement with analytic calculations of quasi-normal modes in the supergravity approximation.

There are several natural extensions to this work. For instance, one might be able to study Lyapunov exponents in classical statistical simulations similar to Refs.~\cite{Kunihiro:2010tg,Gur-Ari:2015rcq,Hanada:2017xrv}, including $1/N$ effects.

Another natural extension would be to consider simulations of Yang-Mills on a 2-dimensional torus rather than a circle, where a much richer phase diagram is expected based on gravity calculations \cite{Dias:2017coo}. Within the present simulation environment based on Ref.~\cite{Hanada:2016qbz}, such change is operationally trivial and just corresponds to extending the number of lattice sites along a second direction.

To conclude, based on our results obtained in this work  we expect that classical statistical simulations of SU($N$) Yang-Mills theory could become a valuable tool to study real-time phenomena in quantum gravity that are otherwise hard to access.

\section{Acknowledgments}

This research was funded by the Department of Energy, award number DE-SC0017905. We would like to thank Francesco Aprile, David Berenstein, Pavel Buividovich, Tom DeGrand, Óscar Dias, Takaaki Ishii, Andreas Sch\"{a}fer and Toby Wiseman for fruitful discussions.

\subsection*{Personal Note by M.H.}

One of the roots of this work is a conversation with Joe Polchinski in November 2013. 
M.~H. explained the lattice simulations of SYM thermodynamics, then Joe said 
it was very impressive but he also wanted to know the real-time dynamics. 
M.~H. said that no generic tool for real-time quantum simulation is known 
and if he could do it he would immediately go to Stockholm rather than chatting with Joe in California.
Then Joe said, with a mild villain smile, `Of course I know :).'

Since then we had several conversations regarding the real-time dynamics of quantum gravitational systems. 
The classical Yang-Mills simulation is one of the options Joe encouraged us to try. 
Perhaps we are still at a very primitive stage, but we hope that we can make a steady progress toward Joe's dream!

\bibliographystyle{JHEP}
\bibliography{lit}

\end{document}